\documentclass[aip,rsi,preprint,graphicx]{revtex4-1} % for review purposes

\usepackage[T1]{fontenc}
\usepackage[ansinew]{inputenc}
\usepackage{graphicx} %%For loading graphic files
\usepackage{amsmath}
\usepackage{latexsym}
\usepackage{amssymb}
\usepackage{color}
\usepackage[normalem]{ulem}
\usepackage{natbib}
\newcommand{\red}{\color{red}}
\newcommand{\blue}{\color{blue}}

\definecolor{linkcolor}{rgb}{0,0,0.6} %hyperlink

\begin{document}

\title{
  {\ \ \ \ \ \  Simultaneous and accurate measurement of the dielectric constant \\  
  \ \ \ \ \ \ \ \ \ \ \ \ \ \ \ \ \ \  at many frequencies spanning a wide range }  
 }
\date{\today}

\author{Roberto P\'{e}rez-Aparicio}
\author{Caroline Crauste-Thibierge}
\email[Corresponding author E-mail: ]{caroline.crauste @ ens-lyon.fr}
\author{Denis Cottinet}
\author{Marius Tanase}
\author{Pascal Metz}
\author{Ludovic Bellon}
\author{Antoine Naert}
\author{Sergio Ciliberto}
\email[Corresponding author E-mail: ]{sergio.ciliberto @ ens-lyon.fr}

\affiliation{Laboratoire de Physique, Ecole Normale Sup\'{e}rieure de Lyon (ENS/CNRS, UMR 5672), 46 all\'{e}e d'Italie, F69007 Lyon, France}

\begin{abstract}
We present an innovative technique which allows the simultaneous measurement of the dielectric constant of a material at many frequencies, spanning a four orders of magnitude range chosen between $10^{-2}$~Hz and $10^{4}$~Hz. The sensitivity and accuracy are comparable to those obtained using standard single frequency techniques.  The technique is based on three  new and  simple features: a)  the precise real time correction  of the amplification of a current  amplifier; b) the specific shape of the excitation signal and its frequency spectrum; and c) the precise synchronization between the  generation   of the excitation signal and the acquisition of the dielectric response signal. 
This technique is useful in the case of
relatively fast  dynamical measurements when the knowledge of the time evolution of the  dielectric constant is needed.
\end{abstract}


\maketitle

\section{Introduction \label{intro}}

Dielectric spectroscopy has provided many informations to the investigation of complex materials, and especially to the field of glass transition phenomenon (glass-forming systems) during the last decades. This technique gives direct access to the polarization of molecular dipoles allowing the study of relaxation processes. It can probe the molecular structure and dynamics of several materials like liquids, polymer composites, colloids, porous materials, or ferroelectric crystals \cite{Kremer2003}. It covers a wide range of frequency, typically from $10^{-2}$ to $10^6$ Hz, going up to $10^{10}$ Hz with coaxial techniques \cite{Kremer2003,Lunkenheimer2000}. Dielectric liquids of small molecules like the glass-forming glycerol \cite{Lunkenheimer1996} or glass-forming polymers \cite{Bur2002,Gomez2001} have been extensively studied. 
 
Dielectric spectroscopy gives direct access to relaxation times\cite{Havriliak1967} like rheology, but it is a rotational relaxation one, not a translational. Nevertheless, it has been showed that on molecular liquids at equilibrium, these two relaxation times have the same temperature dependence\cite{Chang1997}, at least above the glass transition. 
Recent progress in the glass transition were provided by direct measurement of a dynamical growing length scale\cite{CrausteThibierge2010}. This technique has always been improved, e.g., the important work done regarding the electrodes in order to measure charged liquids even at low frequency \cite{Mellor2012,BenIshai2012}. Dielectric spectroscopy has also been coupled to scattering techniques such as X-rays \cite{Sics2000} or neutrons \cite{JimenezRuiz2005}.  

Furthermore, recent theories on aging and out of equilibrium physics predict interesting low frequency properties of the response and fluctuations, either  during the relaxation of the system towards equilibrium after a quench of the control parameter \cite{Peliti,Kob}, or when the system is  driven by external fields \cite{Barrat,Solano}. In order to test these models it is important to measure the response of the system as a function of time after the beginning of the time dependent phenomenon\cite{Bellon,Herisson,Buisson}.  Single frequency measurements will force to repeat the experiment many times for each selected frequency making the experiment extremely long and difficult. In those cases multi-frequency measurements are absolutely necessary.

The standard technique to perform multi-frequency response functions is to drive the system with a white noise signal. The main disadvantage of this approach is to require long averages before reaching reliable response measurements (see Section \ref{procedure}). Indeed, the lower frequencies need long time sampling. 
To overcome this problem  we present an innovative setup that allows simultaneous impedance measurements at several frequencies over a typical range of $10^{-2}-10^4$ Hz with low noise and high sensibility. This has been obtained with a careful choice of the input signal, of the current amplifier, and an appropriate correction, which allow us to measure high capacitances with very small losses  (>$10^{10}~\Omega$ is the typical order of magnitude). This means that our system is particularly useful for the study of relaxation processes in polymer films which have very high resistivity. The length of the measurement time window is imposed only by the smallest frequency that one needs to resolve because all other frequencies are measured simultaneously. 

The paper is organized as follows. In Section \ref{bkg} we give some general, basic background on the dielectric spectroscopy. Section \ref{principle} describes the experimental principle, \ref{setup}  the experimental setup, and \ref{procedure} the measurement procedure. Application of the technique to the study of some polymer films is presented in Section \ref{samples}. We conclude in Section \ref{sumcon}.

%****************************************************
%********** BACKGROUND *****************************
\section{Background \label{bkg}}

Dielectric spectroscopy investigates the frequency dependent dielectric properties of a material \cite{Kremer2003}.  It is based on the interaction of an external field with the electric dipole moment of the sample, often expressed by the complex dielectric permittivity or dielectric constant:
\begin{equation}
\varepsilon(\omega)= \frac{{\rm D}(\omega)}{{\rm E}(\omega)},
\label{eq-permittivity}
\end{equation}
where ${\rm D}(\omega)$ is the dielectric displacement and ${\rm E}(\omega)$ the electric field, both dependent on the angular frequency $\omega$. $\varepsilon(\omega)$ can be expressed as
\begin{equation}
\varepsilon(\omega)= \varepsilon'(\omega)- i\varepsilon''(\omega),
\label{eq-permittivity2}
\end{equation}
being $\varepsilon'(\omega)$ the stored permittivity, which is related to the stored energy by the sample, and $\varepsilon''(\omega)$ the loss permittivity, which indicates the loss energy. Their loss tangent is
\begin{equation}
{\rm tan}(\delta)= \frac{\varepsilon''(\omega)} {\varepsilon'(\omega)}
\label{eq-tandelta}
\end{equation}
where $\delta$ is the phase shift between ${\rm D}(\omega)$ and ${\rm E}(\omega)$. The dielectric constant $\varepsilon$ depends on the dielectric material, the frequency, the temperature, age.

The orientation polarization of dipoles due to applied disturbances
allows the  investigation of  the dielectric relaxation of the material in the frequency range below $10^9~$Hz. Moreover, if those dipoles are of molecular origin, the dielectric spectroscopy allows to investigate molecular motions.
 The experimental techniques and procedures for dielectric spectroscopy depend on the frequency range of interest. At low and intermediate frequencies ($10^{-3} - 10^{7}$ Hz) the sample can be modeled as a parallel plate capacitor where the sample is the dielectric insulator in between the two electrodes \cite{guard}.  
Assuming no edge capacitances, the capacitance of an ideal parallel plate capacitor reads
\begin{equation}
C(\omega) = \frac{\varepsilon_0\varepsilon'(\omega) S}{d}
\label{eq-permittivity1},
\end{equation}
where $d$ is the distance between electrodes, $S$ the surface occupied by the sample and  $\varepsilon_0$ the vacuum permittivity. 
Using Eq. \ref{eq-permittivity2} in Eq. \ref{eq-permittivity1}, we see that the impedance $Z(\omega)$ of the sample  is modeled by a resistor $R(\omega)$ in parallel with a capacitor $C(\omega)$ and it can be expressed as :
\begin{equation}
\frac{1}{Z(\omega)} =\frac{1}{R(\omega)}+i C(\omega)\omega
\label{eq-losses}
\end{equation}
with $
\varepsilon'(\omega)=C(\omega)(\varepsilon_0 S)^{-1}d
$ 
and $\varepsilon ''(\omega)=(\varepsilon_0  S R(\omega)  \omega )^{-1}d$.

The complex impedance $Z(\omega)$ can be measured  by imposing  the voltage $V(\omega) e^{i\omega t}$ and measuring the current $I(\omega) e^{i\omega t} $ flowing through: 
\begin{equation}
Z(\omega)= \frac{V(\omega)}{I(\omega)}
\label{eq-impedance1}
\end{equation}
As $\varepsilon''$ is low on pure dielectric systems this means that the resistive part of the impedance is quite high and the in-phase current becomes hardly detectable.

%********************************************
%************   PRINCIPLE    *******************
\section{The principle of the  measure \label{principle}}

So far we described the ideal capacitor, but in the real experiments there are other parameters to take into account. 
On the one hand, corrections due to the measurement system like the influence of the extra capacitance in the  cables. 
On the other hand, sample preparation must be very careful to avoid difficulties, such as sample geometry, bad electric contacts between electrodes and sample, edge capacitances, or electrode polarization. 
Sample geometry has to be well known. We do not discuss these standard experimental details but we focus on the description of the features of the electronics, the driving signal, and the data analysis that we use in our experimental setup.

\begin{figure}
\centering
\includegraphics{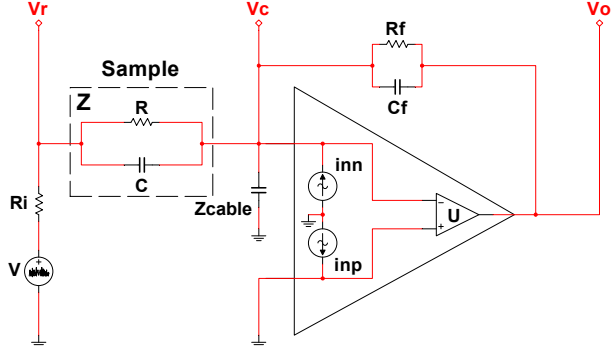}
\caption{\label{fig-circuit}
Scheme of the working principle of the main amplifier of our device.}
\end{figure}

Our goal is to measure $\varepsilon(\omega)$ simultaneously in a frequency range covering 4 orders of magnitude in $\omega$.
As dielectric losses are in general small in pure dielectric (such as polymers),  any spurious effect must be taken into account  in order to have a good accuracy on  the wide frequency range. To define  notation and to clarify the principle of our technique  let us recall basic electronic concepts.   In order to measure $I(\omega)$ we use a current amplifier whose  entrance impedance is just the sample impedance  $Z(\omega)$ as sketched in Fig. \ref{fig-circuit}. In order to analyze the drawbacks of this scheme let
us assume first that the amplifier U in Fig. \ref{fig-circuit} is an ideal one. Thus, we suppose that the bias currents at both entrances are zero
 ($i_{nn}=i_{np}=0$), and that the open-loop amplification of U is infinite for all frequencies.   This implies that $V_c=0$  because 
  the positive entrance is at the ground potential.  Therefore, the  impedance $Z_{cable} \rightarrow \infty$  (given by the parallel of the capacitances of the  cables and of the amplifier entrance) does not play any role in this case. From the standard circuit analysis,  the current balance between the current flowing in the feedback impedance $Z_{f}$ ($C_f // R_f$) and the entrance current  we obtain
\begin{equation}
\frac{V_o}{Z_{f}}=-\frac{V_r}{Z},
\label{eq-ideal}
\end{equation}
where $V_r$ is the reference and $V_o$ the output potentials (see Fig. \ref{fig-circuit}). We have omitted the obvious dependence on $\omega$. Thus, the impedance $Z$ can be calculated as
\begin{equation}
Z=-Z_{f}\frac{V_r}{V_o}
\label{eq-ideal2}
\end{equation}
 
Nevertheless, considering a real amplifier, (i.e. with a finite open loop gain)   one has to take into account in the current balance  
(Fig. \ref{fig-circuit}) that $V_c\ne0$. Thus, $Z_{cable}$ cannot be neglected and the current  balance is  :
\begin{equation}
\frac{V_o-V_c}{Z_{f}}+\frac{V_r-V_c}{Z}-\frac{V_c}{Z_{cable}}+  {\delta  I_o}=0,
\label{eq-real}
\end{equation}
where we have considered that the random output current   ${\delta  I_o}$ takes generically into account all the sources of noise produced by  
 the entrance bias currents, the voltage noise, and the component thermal noise.
In order to get $Z$ we first multiply Eq. \ref{eq-real} by the  conjugate complex $V_r^*(\omega)$ of $V_r(\omega)$ and we take an average 
over several realizations:
\begin{eqnarray}
\frac{<V_oV_r^*>-<V_cV_r^*>}{Z_{f}}&+&\frac{|V_r|^2-<V_cV_r^*>}{Z}  \notag\\
               &-&\frac{<V_cV_r^*>}{Z_{cable}}=0,
\label{eq-real_2}
\end{eqnarray}
where we took into account that  ${\delta  I_o}$ is uncorrelated with $V_r$ thus $<\delta  I_oV_r^*>=0$. 
We consider the transfer functions of both the signal and the correction:
\begin{equation}
G_{sig}(\omega)=\frac{<V_o(\omega)V_r^*(\omega)>}{|V_r|^2(\omega)}
\label{eq-transferinput}
\end{equation}
\begin{equation}
G_{corr}(\omega)=\frac{<V_c(\omega)V_r^*(\omega)>}{|V_r|^2(\omega)}
\label{eq-transfercorr}
\end{equation}

Using Eq. \ref{eq-real_2}, and the definitions of $G_{sig}$ and $G_{corr}$, the impedance of the sample can be calculated as
\begin{equation}
\frac{1}{Z}= \left(\frac{G_{corr}}{Z_{cable}}
+\frac{G_{corr}-G_{sig}}{Z_{f}}\right) \frac{1}{1-G_{corr}}
\label{eq-impedance}
\end{equation}

The knowledge of $V_c$ is very important in order to have reliable estimations of the dielectric 
constant at relatively high frequencies when the amplifier open loop gain begin to decrease and the phase shifts are not negligible 
with respect to the one that we want to measure. In our experimental setup we actually measure $V_c$  as we describe in the next section.

%********************************************
%********* EXPERIMENTAL SETUP ***************
\section{Experimental setup }\label{setup} 

\begin{figure}
\centering
\includegraphics{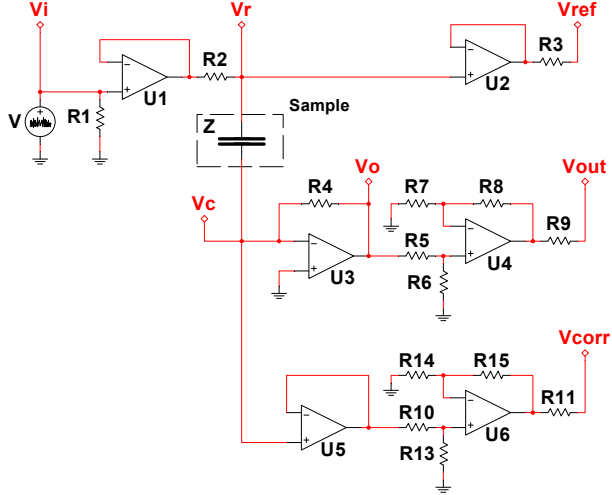}
\caption{\label{circuitall}
Scheme of our real circuit. $\rm{U_1}$, $\rm{U_2}$, and $\rm{U_4}$ are low noise voltage amplifiers. $\rm{U_3}$ and $\rm{U_5}$ are low noise current amplifiers. Details of the components are on Table \ref{table1}.}
\end{figure}

The real scheme of our device is presented in Fig. \ref{circuitall}. The characteristics of each component are presented in Table \ref{table1}. The source signal $V(t)$ is sent to a first operational amplifier $\rm{U_1}$. This amplifier is a voltage follower or a non-inverting buffer. This leads to decouple the impedance of the source and the sample with a negligible additional noise. For that, we choose the LME 49990 amplifier, which has very low voltage noise. The output of $\rm{U_1}$  is sent to the sample via $R_2$ which is the equivalent to $R_i$ in Fig. \ref{fig-circuit}. Note that $R_i=R_2$  is very useful  in order to avoid self oscillations of the operational amplifier $\rm{U_1}$ when large $C$ are used  (see Appendix\ref{app}). 
The signal sent to the sample is  $V_r$ as defined in the previous section. 

\begin{table}[htbp]\centering
\caption{\label{table1} Technical details of the components of the circuit of Fig. \ref{circuitall}. U are the different amplifiers of types LME 49990 (low voltage noise, 1.4~nV/$\sqrt \textrm{Hz}$ at 10~Hz) or AD549 (low  current noise, 0.5$\cdot10^{-15}$A$_{rms}$, $f=~0.1$ to 10~Hz, bias current 30~fA); R are resistances.}
\begin{tabular}[t]{c  c|| c  c}
       \hline
       \hline
        \multicolumn{2}{c||}{U} & \multicolumn{2}{|c}{R ($\Omega$)} \\
       \hline
        $\rm{U_1}$ & LME 49990 & $\rm{R_1}$ & $10^4$ \\
        $\rm{U_2}$ & LME 49990 & $\rm{R_2=R_5=R_7=R_{10}=R_{14}=R_{i}}$ & $10^3$ \\
        $\rm{U_3}$ & AD549 & $\rm{R_3=R_9=R_{11}}$ & $50$ \\
        $\rm{U_4}$ & LME 49990 & $\rm{R_4=R_f}$ & $10^6$ \\
        $\rm{U_5}$ & AD549 & $\rm{R_6=R_8=R_{13}=R_{15}}$ & $4 \cdot 10^3$ \\
       \hline
       \hline    
\end{tabular}
\end{table}

%*******************
%*******************

$V_r$  is also sent to a secondary buffer amplifier $\rm{U_2}$ (LME 49990), and the output of this amplifier is 
used as the reference signal $V_{ref}$ (assuming $V_{ref}=V_r$ because the phase-shift  of $\rm{U_2}$  is negligible in our frequency range).

The current flowing through the sample can be very  small if the sample impedance is very high  so this part of the circuit uses amplifiers
with a low current noise and bias current (AD549). There are two acquisition paths: in the main one, the voltage from the sample goes through a first amplifier $\rm{U_3}$ with $R_4 = R_{f} = 10^6~\Omega$. 
 This amplifier is the main one of the circuit and is equivalent to the one  modeled in Fig. \ref{fig-circuit} and  by Eq. \ref{eq-real}. It is followed by
  $\rm{U_4}$ (LME 49990) which amplifies the voltage by a factor 4 taken into account in data treatment. 
  The second path is the measurement of the negative entrance potential of amplifier $ \rm{U_3}$, that is to say $V_c$ in Eq. \ref{eq-real}. 
This signal also goes through a buffer amplifier $\rm{U_5}$ and is amplified 4 times by $\rm{U_6}$, getting $V_{corr}$ (i.e. $V_{corr}/4= V_c$). 

In order to accurately measure impedances, we have to take into account all the contributions 
from the different components in the electronic device. The impedance of the cables $Z_{cable}$ between the sample and the amplifiers has to be considered
\begin{equation}
Z_{cable}=\frac{1}{i \omega (C_{cable}+C_e)},
\label{eq-impedancecables}
\end{equation}
where $C_{cable}=~50 \cdot 10^{-12}$~F is the capacitance of 0.5 m long cables we used. The entrance capacitance $C_e\simeq 10^{-12}$~F
of $ \rm{U_3}$ can be neglected. We made careful choice of good quality cables between the sample and the amplifier --other cables have less sensitive contribution to the noise in the circuit-- in order to reduce noise in the signal.
We also have to take into account the feedback impedance of the amplifiers
\begin{equation}
Z_{f}=\frac{1}{R_{f}^{-1} + i \omega C_{f}},
\label{eq-impedanceconvert}
\end{equation}
where $R_{f}=~10^6~\Omega$ is the feedback resistance and $C_{f}=~10^{-13}$~F its parallel capacitance (Fig. \ref{fig-circuit}). This capacitance is negligible compared to the impedance of $R_{f}$ in our frequency range.

The signals $V_{ref}$, $V_{corr}$, and $V_{out}$ are sent to a National Instrument PXI-4472 acquisition card (24 bits resolution). 
Data acquisition is performed via a Labview home-made program and data treatment using Matlab to compute the transfer functions $G_{sig}(\omega)$ (Eq. \ref{eq-transferinput}) and $G_{corr}(\omega)$ (Eq. \ref{eq-transfercorr}) using standard FFT algorithm. The transfer functions of the three amplifiers composed by  $\rm{U_2}$,
$\rm{U_4}$, and ($\rm{U_2}$+$\rm{U_6}$) are measured in order to correct the phase of the three measured signals.

%********************************************
%***********    PROCEDURE   ******************
\section{The measurement procedure}
\label{procedure}

In order to  perform a simultaneous and precise measurement of $Z$ in a wide frequency range using the device described in the previous section we have optimized the frequency dependence of the  signal $V_r$. Indeed the standard technique employed to perform a response measurement on a wide band is to excite the system with an almost white noise in the band of interest and to measure simultaneously the input and output signals.  Then, by taking FFT of these two signals input and output, one can compute the transfer function $G_{sig}$ by a very long average of the response function on many configurations of the noise. This system although very practical and adaptable on almost all cases presents several drawbacks in the case of a circuit like the one presented in the previous sections. The first is that in the circuit of Fig. \ref{circuitall} the high frequencies are amplified much more than the low ones. Therefore, one has to compensate for this effect. The second drawback is using a random noise at the input, its  phase is rapidly changing and many averages are required in order to smooth this phase noise and to have a reliable measurement. In order to avoid these  difficulties and reduce the number of averages in our system we apply the following method.

%********************************************
%***********    Method  ******************
\subsection{Method}
\label{method}
\begin{figure}[htbp]
\begin{center}
\includegraphics{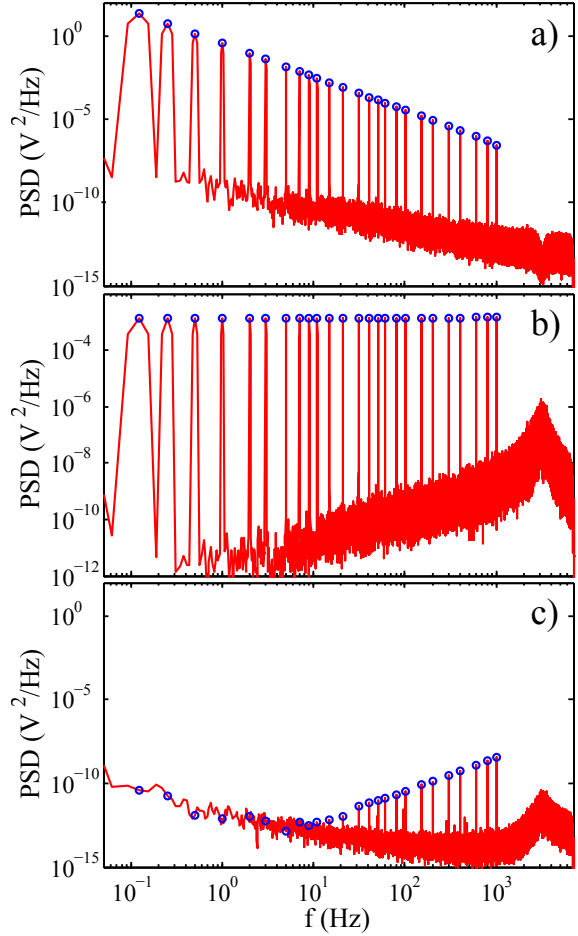}
\end{center}
\caption{Power spectral density of (a) $V_{ref}$, (b) $V_{out}$, and (c) $V_{corr}$ signals. Circles point to the chosen frequencies.}
\label{fig-psdfreq}
\end{figure}

We first fix a measurement interval $T$ and a sampling frequency $f_s= N/T$, where $N$ is an integer.  Then, we generate a signal composed by a sum of sinusoids whose pulsations $\omega_k$ are harmonics of the slowest frequency $2\pi/T$, specifically 
\begin{equation}
V(t)= \sum_k V_{0,k} \sin{\omega_k t},
\label{eq-sin}
\end{equation}
where $V_{0,k}$ are the amplitudes in volts. The amplitudes are chosen to be $V_{0,k}\propto1/\omega_k$ because favoring low frequencies allows us  to compensate the increasing gain of the amplifier at high frequencies. These kinds of signal are  numerically generated and  recorded in the computer memory which is then sampled by { the same clock} at the sampling frequency $f_s$, 
which is also used for sampling the acquired signals $V_r$, $V_o$, and $V_c$. In such a way all phase noises are suppressed as each frequency contains a fix  integer number of acquired samples.   
We use two different devices to generate the signal either  a National Instrument PXI 5411-card generator or an Agilent 33500B Series waveform generator. 
As an example we plot  in Fig. \ref{fig-psdfreq} the spectra of $V_{ref}$, $V_{out}$, and $V_{corr}$ signals obtained using  $C=10^{-8}$~F.  
In this specific example we select twenty-five frequencies $f_k~=~\omega_k/(2\pi)$  in a logarithmic scale between a minimum value of $0.125$~Hz and  $10^3$~Hz. The sampling rate is  $2^{15}=32768$ sample/s. 
The peak of the noise  at about 3~kHz is explained in the Appendix\ref{app}. Notice that the amplitude of the correction signal is not negligible above $100$~Hz (Fig. \ref{fig-psdfreq}c) and it is important to keep
 it into account in the calculation of $Z$. To summarize, this electronics combined with the described data analysis can measure with a good accuracy $R\leq10^{12}~\Omega$ and $C>10^{-11}$~F.

%*******************
%*******************

\begin{figure}[htbp]
\begin{center}
\includegraphics[width=\textwidth]{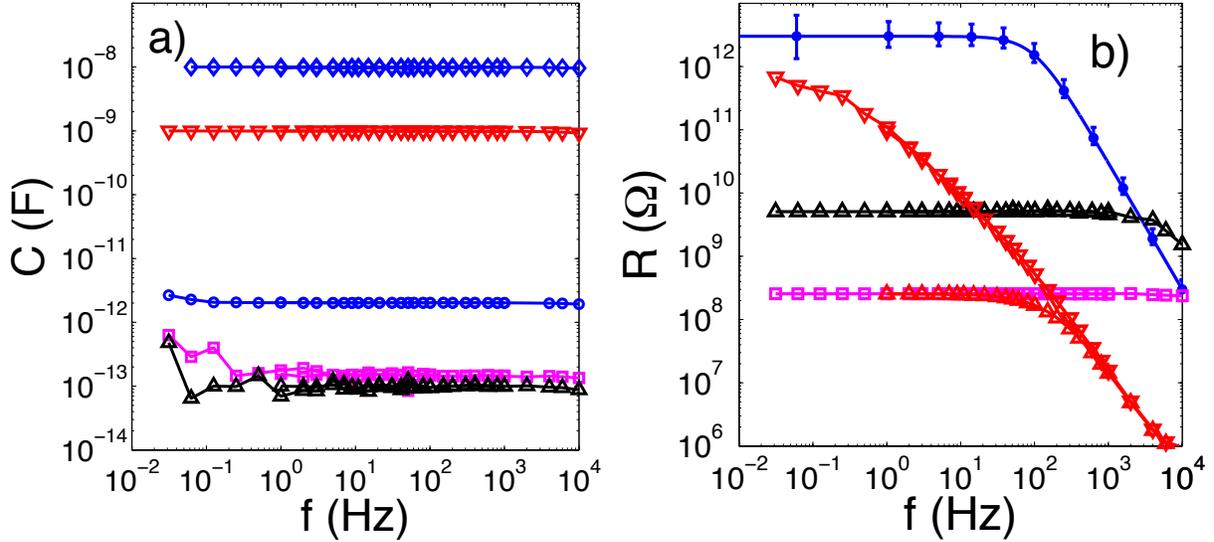} 
\end{center}
\caption{{ Capacitances  (a) and resistances  (b) of several components used to check the accuracy and the limits of our apparatus. 
Three  capacitances of $2.5$ pF({\blue$\bullet$}), $1$ nF({\red$\bigtriangledown$}) and $10$ nF ({\blue$\diamond$}) are plotted. The losses of the  $2.5$ pF({\blue$\bullet$} with error bars) and 
 $1$nF{\red$(\bigtriangledown)$} capacitances are plotted in (b), whereas those of the $10$ nF in fig.\ref{fig-capatest}b.   The capacitance of $2.5$ pF has very small losses and this allows us to fix the limits and the losses  of our apparatus (blue line with error bars in b). 
In (b) we also plot the value of two pure resistances of $250$ M$\Omega$ ({\color{magenta}$\square$})and $5$~G$\Omega(\bigtriangleup)$. These two resistances have of course a small parallel capacitance of about $0.1$ pF shown in (a) {\color{magenta}$(\square)$} and $(\bigtriangleup)$, which also fixes  the accuracy of our system. In (b) we also plot the measured resistance  {\red($\bigtriangleup$)} when $Z$ is the 
 parallel of the  $1$nF capacitance with the $250$ M$\Omega$ resistance. Notice that for each $Z$ the whole frequency interval is covered by only two simultaneous measurements in two overlapping four order of magnitude intervals: [$10^{-2}-10^2$ Hz] and [$1-10^4$ Hz]  }} 
\label{fig_all_Z}
\end{figure}

\begin{figure}[htbp]
\begin{center}
\includegraphics[width=\textwidth]{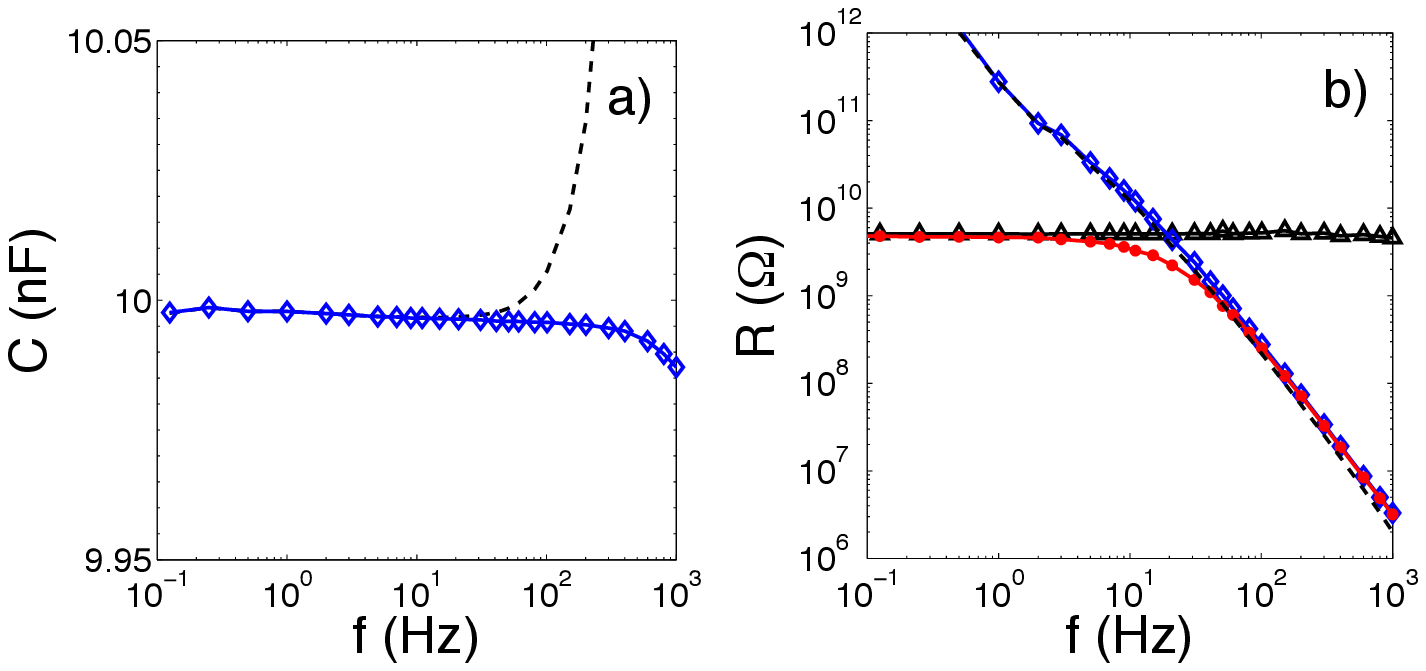}
\end{center}
\caption{{
(a)  Expanded view of the measured $10$ nF capacitance of Fig.\ref{fig_all_Z}a. Its losses are  plotted in (b{\blue$\diamond$}) together with the pure resistance of $5$ G$\Omega(\bigtriangleup)$. We also plot the measured  total resistance {\red$\bullet$}  of $Z$ composed by the parallel of the $10$ nF with the $5$ G$\Omega$ resistance.  The dashed line represent the results of the measurement  without using the correction signal $V_c$ in the calculation. The whole measurement in the range 0.1 Hz-1000 Hz takes less than 60~s (i.e. 6 periods of the smallest frequency).} }
\label{fig-capatest}
\end{figure}

%********************************************
%***********    Data Analysis  ******************
\subsection{Data Analysis}
\label{analysis}

From the measured impedance (Eq. \ref{eq-impedance}) we can calculate the resistance and the capacitance of the dielectric as
\begin{equation}
R=\left[ Re \left(\frac{1}{Z}\right)\right]^{-1}
\label{eq-resistance}
\end{equation}
\begin{equation}
C=Im \left(\frac{1}{Z}\right) \omega^{-1}
\label{eq-capacitance}
\end{equation}
In order to check the efficiency of our device in the chosen frequency window,  we use several  impedances $Z$  whose values have been checked  by standard apparatuses \cite{Note1}. In Fig.\ref{fig_all_Z}a we plot the measured values of the capacitances and in Fig.\ref{fig_all_Z}
b) those of the measured resistances. This plot illustrates the whole dynamics of our system ranging from $1$pF (at $0.1$pF the error is large but the measurement  can be still used)  to $10$ nF and larger if a smaller $R_f$ is used \cite{Note2}. The $2.5$ pF capacitance is an air capacitance with extremely small losses and this allows us to determine the losses of our circuit as a function of frequency which are plotted in Fig.\ref{fig_all_Z}b. In order to appreciate more precisely the accuracy we plot in Fig.\ref{fig-capatest}a) 
the measured values of the $10$ nF capacitance and Fig.\ref{fig-capatest}b  its losses in the frequency range  $0.1\rm{~Hz}-1\rm{~kHz}$. We see that the fluctuations of the measurement are less  than $1/1000$ in all the frequency range. In Fig.\ref{fig-capatest}b we compare the losses of the $10$ nF capacitance with a measured resistance $R_e=5$ G$\Omega$.
The capacitance $C$ has an almost constant value, whereas its  losses increase, which correspond  to a decrease of $R(\omega)$ as a function of $\omega$ (see eq.\ref{eq-losses} and Fig.\ref{fig-capatest}b). When  $C$ is mounted in parallel to $R_e$  then the capacitance $C$ is unchanged whereas the measured resistance is now the parallel of $R_e//R(\omega)$ which is constant at low frequencies up to around 10 Hz, and then, decreases as a function of the increasing frequency due to the capacitive losses \cite{Note1}. It is important to stress that, using our method, we need to average less than $60$~s, in order to  measure simultaneously the values of $R$ and $C$  on 4 orders of magnitude in frequency, plotted in Fig. \ref{fig-capatest}. As already mentioned, standard techniques use a white noise in the bandwidth of interest. By applying such a signal to  our circuit, the result is much more noisy (even impossible to find correct values) than that obtained using our specifically chosen input signal described in the previous section (see Eq.\ref{eq-sin}).
In Fig. \ref{fig-capatest} we can also see the importance of the correction signal $V_c$ ( see Fig. \ref{fig-psdfreq}c) in order to properly calculate the values of $R$ and $C$ in the whole chosen bandwidth. { Finally we stress that  { because of the limited memory of our voltage generator we always measure simultaneously at most 4 order of magnitude in frequency in the
range $0.01\rm{~Hz}-10\rm{~kHz}$. Furthermore the accuracy is not constant as a function of frequency. It is $5~\%$ below $0.1$ Hz and above $1$ kHz about $1~\%$ in the range $0.1\rm{~Hz}-1\rm{~kHz}$}}

%**********************************************************
%***************** APPLICATION ****************************
\section{Materials and Applications \label{samples}}

In the case of a pure capacitance, from the values  of $R$ and $C$ we can  calculate both $\varepsilon'$ and $\varepsilon''$:

\begin{equation}
\varepsilon'=\frac{C}{C_0}
\label{eq-epsilon1}
\end{equation}
\begin{equation}
\varepsilon''=\frac{1}{R \omega C_0}
\label{eq-epsilon2}
\end{equation}
with $C_0=S \varepsilon_0 d^{-1}$. Therefore, ${\rm tan} (\delta)$ (Eq. \ref{eq-tandelta}) can be calculated as
\begin{equation}
{\rm tan} (\delta)=\frac{1}{R C \omega},
\label{eq-tandelta2}
\end{equation}
which is independent of the geometry of the sample.

\begin{figure}[htbp]
\begin{center}
\includegraphics{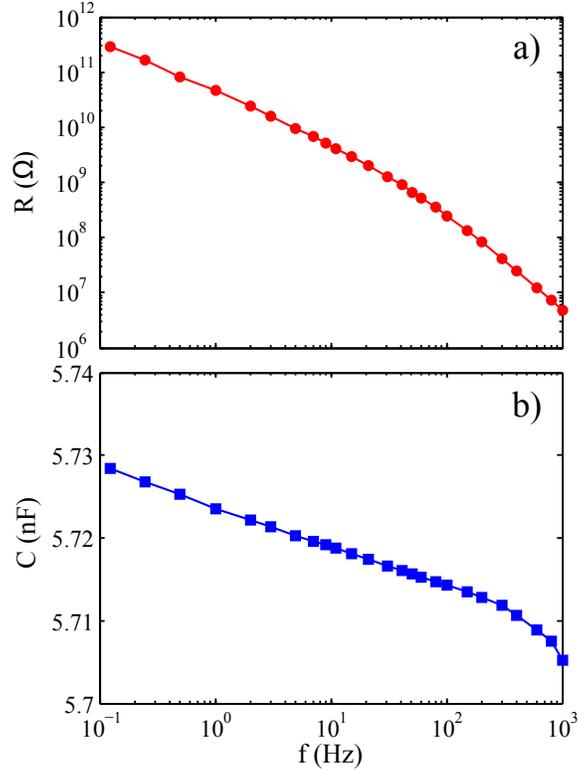}
\end{center}
\caption{Resistance (a) and capacitance (b) of a PC film of $125~\mu$m of thickness and $18$~cm of diameter as a function of frequency at room temperature. }
\label{fig-pc}
\end{figure}

\begin{figure}[htbp]
\begin{center}
\includegraphics{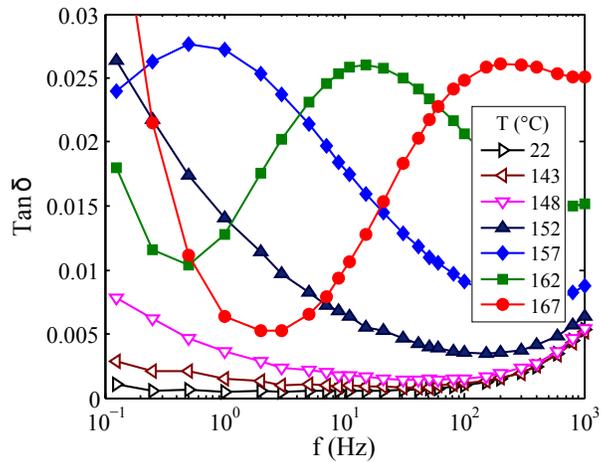}
\end{center}
\caption{Tan~$\delta$ of a PC film of 125~$\mu$m of thickness and 18~cm of diameter as a function of frequency at different temperatures.}
\label{fig-pctandelta}
\end{figure}

We check our experimental device with well-known polymer films. As already mentioned, our experimental device allows to perform simultaneous measurement of impedance at several frequencies in a range from $10^{-2}$ to $10^3$ Hz with low noise and high sensitivity. 
First of all, we investigate an extrusion film MAKROFOL\textsuperscript{\textregistered} DE 1-1 000000 (from BAYER) based on Makrolon\textsuperscript{\textregistered} polycarbonate (PC) of 125~$\mu$m of thickness and with a glass transition temperature ($T_g$) of about 150~$^\circ$C. The sample consist of a PC film of 18~cm of diameter. 
We measure the impedance and compute $C$ and $R$ from 22 to 167~$^\circ$C. Fig. \ref{fig-pc} shows $R$ and $C$ at 22~$^\circ$C. $R$ presents huge values at low frequencies decreasing with frequency. 
$C$ presents a small dependence in frequency. 
From $C$ we calculate the dielectric constant $\varepsilon$ (Eq. \ref{eq-permittivity1}) to be similar to the value found in the literature\cite{Mark1999} ($\varepsilon(60~\rm{Hz})= 3.17$). 
${\rm tan}~\delta$ (Eq. \ref{eq-tandelta2}) as a function of temperature is presented in Fig. \ref{fig-pctandelta}.
A peak corresponding to the segmental relaxation of PC (glass transition) appears from low to high frequencies when increasing temperature (well-known in polymers, see e.g. ref. \citenum{Lunkenheimer2000}).
The advantage here is that one can do easily multi-frequency dynamical measurements changing the rate of the  temperature ramp. 

%*******************
%*******************

%*******************
%*******************

%*******************
\begin{figure}[htbp]
\begin{center}
\includegraphics{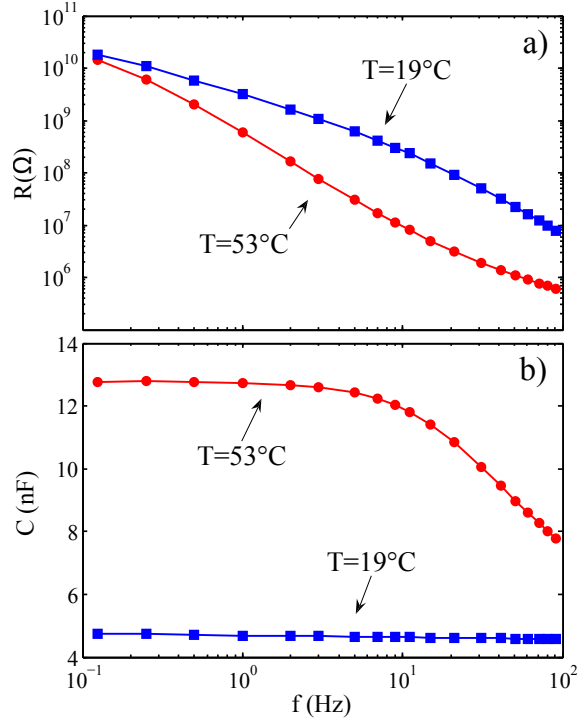}
\end{center}
\caption{Resistance (a) and capacitance (b) of a PVAc sample as a function of frequency at two different temperatures.}
\label{fig-pvacr}
\end{figure}
%*******************

%*******************
\begin{figure}[htbp]
\begin{center}
\includegraphics{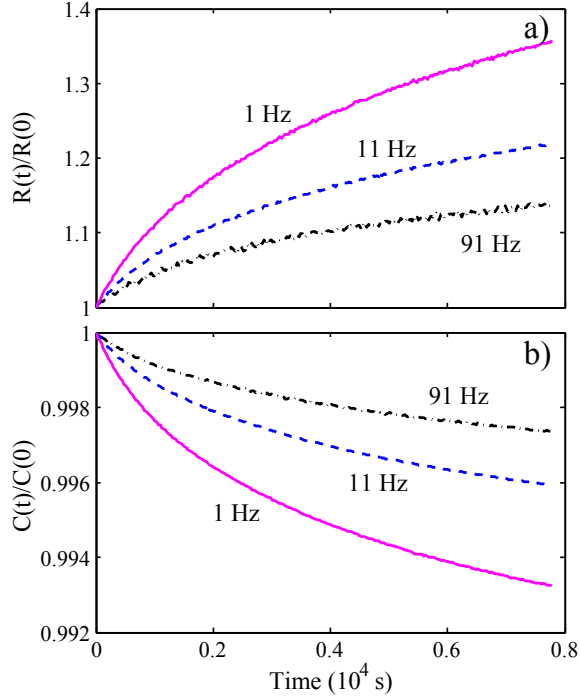}
\end{center}
\caption{Resistance (a) and capacitance (b) of a PVAc sample as a function of the aging time after a quench from 53 to 19~$^\circ$C at 1, 11, and 91 Hz. Both magnitudes are normalized to their values just after the quench (time~$=0$): $1.95 \cdot 10^{9}~\Omega$ (1~Hz), $2.5 \cdot 10^{8}~\Omega $ (11~Hz), and $3.4 \cdot 10^{7}~\Omega$ (91~Hz) for the resistance; $5.99 \cdot 10^{-9}$~F (1~Hz), $5.89 \cdot 10^{-9}$~F (11~Hz), and $5.82 \cdot 10^{-9}$~F (91~Hz) for the capacitance.}
\label{fig-pvacr2}
\end{figure}
%*******************

Then we investigate a film of poly(vinyl-acetate) (PVAc) in order to study the aging phenomenon. 
The PVAc ($T_g \sim 34^\circ$C) is cycled between 19 and 53$^\circ$C, with a very fast thermal quench (around 30$^\circ$C/min).
The electrodes are made out of thin films of aluminum (150~nm) deposited by evaporation over a sapphire disc. 
Such crystal plate is chosen for its high thermal conductivity because heating and cooling are done through the substrate. 
The polymer is dissolved in chloroform, then poured over the lower electrode. After careful drying of the solvent, the second aluminum thin film is deposited over the dielectric. Connexions are fixed on the electrodes thanks to conductive glue. The thickness of the dielectric is around $1~\mu m$, such that the overall capacitance is a few nF in order to match the amplifier's impedance. 
Fig. \ref{fig-pvacr} shows resistance and capacitance of PVAc above and below $T_g$ as a function of frequency. 
The $\alpha$-relaxation appears in the measurement range around 53$^\circ$C but not at lower temperature (19$^\circ$C). Fig. \ref{fig-pvacr2} shows the time evolution of $R$ and $C$ after a fast thermal quench from 53$^\circ$C  to 19$^\circ$C at 1, 11, and 91 Hz.
The relaxation  of $R$ and $C$ are due to the aging of the polymer below the glass transition going toward equilibrium. 
This technique allows to work on a time range limited at small times by the speed of the quench (typically one minute). Our device is absolutely necessary in this kind of dynamical studies.

%******************************************
%****   SUMMARY AND CONCLUSIONS   ********

\section{Conclusion \label{sumcon}}

We present here a technique for impedance measurements, specially designed for highly resistive samples and time dependent phenomena. Indeed it allows us to measure the complex dielectric constant simultaneously in a  wide frequency range. The electronics combined with the described data analysis allow to measure with a good accuracy $R\leq10^{12}~\Omega$ and $C>10^{-11}$~F.

This technique is based on two main elements. First of all, a home-made amplifier specially built to measure small phase shifts in the whole frequency range of interest. The proposed method based on a correction signal is very efficient for obtaining  accurate values of both on the real and the imaginary part of the impedance, even in the cases of very small phase shift. 
  
Secondly, the reference signal is carefully chosen to give optimal accuracy over the whole frequency range, which is typically $10^{-2}-10^4$ Hz. We use a sum of sinus with decreasing intensity at high frequency in order to keep a constant signal-to-noise ratio with frequency.  The chosen frequencies of the input signal are exact subharmonic of the sampling frequency, which is used to generate 
the input signal and to  sample the outputs.  
Compared to the use of a white noise, this signal has well-defined phases, ensuring the good phase accuracy we need.
 Furthermore, measuring known frequencies allows shorter acquisition times than using white noise in order to get high precision on the impedances. Then, this device is particularly useful for time evolving impedances studies, especially in the field of aging of glass-forming polymers. Four orders of magnitude in frequency, chosen in the range of $10^{-2}-10^4$ Hz, can be simultaneously investigated over typical time range from one minute to arbitrary long time.   
We also present here few results on two polymer samples: a PC film evolving to the glass transition with temperature, and a PVAc sample evolving with time after a fast quench. This last result demonstrates  the  rich field of non-equilibrium physics that can be probed using this  device.

%******************************************
%*********   ACKNOWLEDGMENTS   ***********
\section*{Acknowledgments}

R. P\'{e}rez-Aparicio acknowledges the funding by Solvay. Caroline Crauste-Thibierge and Sergio Ciliberto thank the founding by ERC grant 267687 OutEFLUCOP. We also thank Debjani Bagchi who made use of the setup presented here.

%******************************************
%*************   APPENDIX   *****************
\appendix*
\section{ Response to noise }\label{app}

In this appendix we give an explanation of the noise peak in the output spectrum of Fig. \ref{fig-psdfreq}. 
The amplification of the AD549 is very well approximated at low frequencies by
\begin{equation}
A=\frac{A_o } {1+ i \ \omega/ \omega_c }
\label{eq_ampli_response}
\end{equation}
with $A_o\simeq 10^6$ and $\omega_c \simeq 2 \ \pi \ \rm{rad/s}$.

\begin{figure}[htbp]
\begin{center}
\includegraphics[width=8cm]{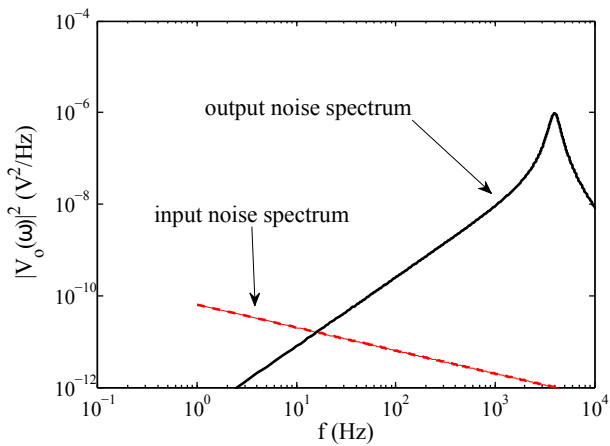}
\end{center}
\caption{Output spectrum (line) computed using in  Eq. \ref{eq_power_spectrum} the  plotted input  spectrum (dashed-line).}
\label{fig-simul}
\end{figure}

The analysis of the circuit of Fig. \ref{circuitall} shows that the spectrum of the output signal $|V_o(\omega)|^2$ is related to that at the input  $|V_i(\omega)|^2$ :
\begin{equation}
|V_o(\omega)|^2=\frac{ (R_f\  \omega \ C)^2  \  A_o^2  \   |V_i(\omega)|^2 } {D(\omega)}
\label{eq_power_spectrum}
\end{equation}
where 
\begin{equation}
\begin{split}
D(\omega)= (1- \omega^2(R_i+R_f) C/ \omega_c +A_o)^2 + \\ 
 \left[( R_i+R_f) \ C+1/\omega_c+A_o R_i  C \right]^2 \omega^2
\label{eq_denominator}
\end{split}
\end{equation}
Here we have used the approximation $R\rightarrow \infty$ and $C_f\simeq 0$ (see Fig. \ref{fig-circuit}).
These are very good for the purpose of understanding the maximum of the noise. 
This spectrum has a resonance   at $\omega \simeq \sqrt{A_o \ \omega_c/[(R_i+R_f) C]}$ which is rather broad because of the strong damping  term $A_o R_i  C $.  Inserting the numerical value we see that the resonance is at about $3.9~\rm{kHz}$ and the broadening a few Hz. These values  are in agreement with the  enhancement of the noise in the power spectrum of Fig. \ref{fig-psdfreq}b, 
if we consider that the noise of the input signal, Fig. \ref{fig-psdfreq}a, has a  wide bandwidth with a slow decrease as function of frequency. 
In Fig. \ref{fig-simul} we plot an  output spectrum  computed from Eq. \ref{eq_power_spectrum} using a realistic input spectrum, which is plotted in the same figure. We see that the output spectrum is very close to the one measured with $C=10^{-8}~$F. 

%******************************************
%************   BIBLIOGRAPHY   **************

%\bibliography{DStechnique_v8}

\begin{thebibliography}{22}
\bibitem [{gua()}]{guard}%
  \BibitemOpen
  \bibinfo {note} {We do not discuss here this obvious problem because it is
  not the subject of the paper and it can be easily taken into account in our
  experimental setup by a guard ring.}\BibitemShut {Stop}%
\end{thebibliography}

\providecommand{\noopsort}[1]{}\providecommand{\singleletter}[1]{#1}%

\end{document}